\documentstyle[]{article}
\def\bea{\begin{eqnarray}}
\def\eea{\end{eqnarray}}

\def\beq{\begin{equation}}
\def\eeq{\end{equation}}
\def\ba{\beq\new\begin{array}{c}}
\def\ea{\end{array}\eeq}
\def\be{\ba}
\def\ee{\ea}

\makeatletter
\newdimen\normalarrayskip              
\newdimen\minarrayskip                 
\normalarrayskip\baselineskip
\minarrayskip\jot
\newif\ifold             \oldtrue            \def\new{\oldfalse}
\def\arraymode{\ifold\relax\else\displaystyle\fi} 
\def\eqnumphantom{\phantom{(\theequation)}}     
\def\@arrayskip{\ifold\baselineskip\z@\lineskip\z@
     \else
     \baselineskip\minarrayskip\lineskip2\minarrayskip\fi}
\def\@arrayclassz{\ifcase \@lastchclass \@acolampacol \or
\@ampacol \or \or \or \@addamp \or
   \@acolampacol \or \@firstampfalse \@acol \fi
\edef\@preamble{\@preamble
  \ifcase \@chnum
     \hfil$\relax\arraymode\@sharp$\hfil
     \or $\relax\arraymode\@sharp$\hfil
     \or \hfil$\relax\arraymode\@sharp$\fi}}
\def\@array[#1]#2{\setbox\@arstrutbox=\hbox{\vrule
     height\arraystretch \ht\strutbox
     depth\arraystretch \dp\strutbox
     width\z@}\@mkpream{#2}\edef\@preamble{\halign
\noexpand\@halignto
\bgroup \tabskip\z@ \@arstrut \@preamble \tabskip\z@ \cr}%
\let\@startpbox\@@startpbox \let\@endpbox\@@endpbox
  \if #1t\vtop \else \if#1b\vbox \else \vcenter \fi\fi
  \bgroup \let\par\relax
  \let\@sharp##\let\protect\relax
  \@arrayskip\@preamble}
\def\eqnarray{\stepcounter{equation}%
              \let\@currentlabel=\theequation
              \global\@eqnswtrue
              \global\@eqcnt\z@
              \tabskip\@centering
              \let\\=\@eqncr
              $$%
 \halign to \displaywidth\bgroup
    \eqnumphantom\@eqnsel\hskip\@centering
    $\displaystyle \tabskip\z@ {##}$%
    \global\@eqcnt\@ne \hskip 2\arraycolsep
         $\displaystyle\arraymode{##}$\hfil
    \global\@eqcnt\tw@ \hskip 2\arraycolsep
         $\displaystyle\tabskip\z@{##}$\hfil
         \tabskip\@centering
    &{##}\tabskip\z@\cr}
\newfont{\hr}{msbm10}
\newfont{\ams}{msam10}

\begin{document}

\begin{titlepage}
\setcounter{footnote}0
\begin{center}
\hfill ITEP/TH-14/96\\

\hfill{\it hepth/xxxxmmm}\\
\vspace{0.3in}
{\LARGE\bf Peierls model and vacuum structure in the N=2
supersymmetric field theories}

\bigskip {\Large A.Gorsky \footnote{E-mail
address: gorsky@vxitep.itep.ru, sasha@rhea.teorfys.uu.se}
$^{\dag}$}
\\
\bigskip
$\phantom{gh}^{\dag}${\it ITEP, Moscow, 117 259, Russia}
\end{center}
\bigskip

\begin{abstract}
We suggest the quasiparticle picture behind the integrable structure of
N=2 SYM theory,which arises if the Lax operator is considered as a Hamiltonian
for the fermionic system.We compare the meaning of BPS states with the one
coming from the D-brane interpretation and give some evidence for the
compositeness of the selfdual strings.The temperature
phase transition with the disappearence of the mass gap is conjectured.
\end{abstract}

\end{titlepage}

\newpage
\setcounter{footnote}0

\section{Introduction}

The impressive progress has been achieved in the description of the vacuum
structure of the both field and string theories since the pioneering works of
Seiberg and Witten \cite{sw}.It turned out that the
natural symmetry arguments
fix the data necessary to describe the vacuum structure  to be the peculiar
Riemann surface and one form on it.These data provide the prepotential for
the low-energy effective theory and the exact mass spectrum.However the
dynamical mechanism responsible for the vacuum configuration is still
obscure.

The same data allow the natural interpretation in terms of the integrable
systems intimately related with these Riemann surfaces.The integrable
system serving the pure N=2 SYM theory appeares to be periodical Toda
chain \cite{toda},SYM with the
adjoint matter- Calogero-Moser system \cite{cal}
and SYM with the fundamental matter -integrable spin chains
\cite{chains}.Prepotentials appeared to be the logariphms of the
quasiclassical $\tau$-functions and one form is identified with the action
variable for the corresponding integrable system.The reviews of the approach
can be found in \cite{mm}.Nevertheless in spite of the evident progress in
the conceptual explanations of the objects which seemed to be misterious
previously the dynamical derivation of the  relevant integrable systems is
absent.Therefore any complementary viewpoint on the problem at
hands can shed new light on the vacuum dynamics.

In this letter we attempt to reformulate the problem in terms of the Peierls
model known in the solid state physics .It describes the
dynamics of the 1d fermions on the fluctuating  lattice.Fortunately
some exact results are known for this model both for the continuum \cite{cp}
and  discrete cases \cite{dp} and the Riemann surface
(vacuum state in SYM theory) plays the role of the dispersion law for the
quasiparticle.We consider the specific interaction in the
Peierls model allowing the explicit comparison with the Toda chain.
We also will give interpretation of 4d BPS states in the
quasiparticle terms.With this interpretation the stringy picture  of the BPS
states \cite {vafa} gets interpretation as a "string of the quasiparticles"
filling the forbidden or allowed zones.The phase transition known in the
Peierls model implies the existence of the same phenomena in 4d case
between the phases with  $\Lambda_{QCD}=0$ and $\Lambda_{QCD}\neq 0$.

\section{Peierls model}

In this section we review  the main facts about the Peierls model relevant
for the further consideration.Initially it was formulated to describe the
selfconsistent behaviour of 1d fermions interacting with the fluctuating
lattice and was applied to the analysis of 1d superconductivity.The Coulomn
interaction between the fermions is neglected,the fermions are assumed to be
in the external field defined by the lattice state while the lattice dynamics
itself is modified by the fermions.In what follows we will
discuss both the continuum and discrete Peierls models.

Hamiltonian density for the  simplest continuum model looks as follows
\be
H_{con}=\Psi^{+}\sigma_{3}\partial_{x}\Psi +\Psi^{+}(\sigma_{-}\Delta^{*}-
\sigma_{+}\Delta)\Psi +\Delta^{2}
\ee
where $\Delta $ represents the lattice potential and in general situation
we have the sum $\sum_{n} \kappa_{n} I_{n}$ with KdV potentials $I_{n}$
instead of the single quadratic term.The components of the fermionic wave
function $\Psi=(u,v)$     obey the equations
\be
\partial u_{E}-(\Delta^{2}+\partial\Delta)u_{E}=E^{2}u_{E} \\ \partial
v_{E}-(\Delta^{2}+\partial\Delta)v_{E}=E^{2}v_{E}
\ee
For the discrete
version one has   \cite{dp}
\be
H_{dis}=\sum_{n}
\Psi_{n}^{+}v_{n}\Psi_{n}+\Psi_{n}^{+}c_{n}\Psi_{n+1}+
\Psi_{n}^{+}c_{n-1}\Psi_{n-1} +\sum_{i}\kappa_{i}I_{i}\\
c_{n}=exp(x_{n+1}-x_{n}) \\
I_{0}N=\sum_{n} lnc_{n} ;I_{2}N=\sum_{n}(c_{n}^{2}+v_{n}^{2})
\ee
where $x_{n}$ are the lattice variables.

The first question to be addressed to the model is about its ground state.To
obtain the ground state we have to minimize the Hamiltonian with respect to
the fermionic and lattice variables.The variation over the fermionic
variables results in the Lax equation for the Toda chain system
\be
c_{n}\Psi_{n+i}+c_{n-1}\Psi_{n-1}+v_{n}\Psi_{n}=E\Psi_{n}
\ee
Variation over the lattice degrees of freedom gives rise to the system of the
finite number of algebraic equations.These equations provide the explicit
form of the Riemann surface which corresponds to the solution.In the
simplest case of two first Toda Hamiltonians one has the following system
\cite{dp}
\be
\kappa_{2}=\frac{i}{2\pi}\oint_{e_{1}}^{\mu} \frac{dE}{\sqrt{R(E)}}\\
\oint_{e_{1}}^{\mu} \frac{(2E_{1}-s_{1})dE}{\sqrt{R(E)}}=0\\
\kappa_{0}=\frac{i}{2\pi}\oint_{e_{1}}^{\mu}(E^{2}-\frac{s_{1}E}{2}
+\frac{s_{2}}{2}-\frac{s_{2}^{2}}{8})\frac{dE}{\sqrt{R(E)}}\\
\rho=\frac{i}{2\pi}\oint_{e_{1}}^{\mu}(E+b) \frac{dE}{\sqrt{R(E)}}\\
R=\prod_{i=1}^{4}(E-e_{i});s_{1}=\sum e_{i} ;s_{2}=\sum_{i<j}e_{i}e_{j}
\ee
where b is defined from the proper normalization conditions and
$\mu$ is the chemical potential.Let us note that the system above is the
analogue of the Virasoro constraints in the matrix models.

It is also possible to write down the explicit solutions for the
fermionic wave functions $\Psi_{n}$ and the lattice variables $x_{n}$
in terms of the elliptic functions
\be
\Psi_{n}(z)=b_{n}\frac{\sigma^{n}(z+z_{0})\sigma(z-\alpha_{n})}
{\sigma^{n}(z-z_{0})\sigma(z-\alpha)}
\ee
where $z_{0}$ is the position of the pole of the elliptic function
E(z) defined by
\be
z=\oint_{e_{1}}^{E} \frac{dE}{\sqrt{R(E)}}\\
\alpha_{n}=2nz_{0}+\alpha
\ee
\be
x_{n}-na=ln\frac{\theta_{4}(\rho(n-\nu-\frac{1}{2})}
{\theta_{4}(\rho(n-\nu+\frac{1}{2})}
\ee
and $\nu$ reflects the degeneration of the ground state,$\rho$ is the
fermionic density.The relation between the modulus of the curve and the
parameters of the model looks as follows
\be
\mid \tau \mid \frac{\theta_{2}\theta_{3}\theta_{4}}
{\theta_{1}(\frac{\rho}{2})}=exp(-a)
\ee
where a is the average distance between the lattice cites.

The key feature of the solution
is the appearence  of the fermionic mass gap  which
substitutes the $\Lambda_{QCD}$.It can be also proved that the chemical
potential for the fermions lies in the forbidden zone for the
quasiparticles.To define the dispersion law for the
quasiparticles  we can consider the  periodicity
property of the fermionic wave function on the lattice
\be
\Psi_{n+N}(E)=e^{iNp(E)}\Psi_{n}(E) .
\ee

The dependence of the
quasimomentum p on the energy E provides the dispersion law which can be
represented as a two dimensional Riemann surface $\Sigma$ in 4d space (p,E)
with complex p and E.Fermionic wave function is uniquely defined on this
surface and the number of its zeros coincides with the genus of the curve.

Another important characteristic is the spectrum of the exitations.It can be
divided in two classes.The first one corresponds to the exitations of the
lattice and can be called phonon.Another gapless exitation corresponds to
the charge density wave.
The second type is fermionic one and strongly depends on the fermionic
density.At large density  one has the polyaron type state when
fermions are localized on the  configurations (in the continuum case)
\be
u(x)=const-\frac{2\chi}{ch^{2}(\sqrt{\chi}x)}
\ee

In the opposite limit of the small $\rho$ one has
the delocalized fermionic state
and the lattice potential
\be
u(x)=const+\chi cos(2\sqrt{\chi}x+\phi)
\ee
where $\chi$ is some constant.
It was also shown \cite{cp} that another important states can be discovered
if the external magnetic field is applied.It turns out that these states
with the magnetic quantum numbers have the localized energy levels
in the forbidden zone.

Exact
solution to the ground state implies the possibility to determine the
temperature dependence of the mass gap.Indeed it was shown \cite{cp}
that the mass gap
for the fermions gets renormalized and disappears at some critical value
$T_{c}$.Being translated to the form of the dispersion law it tells us that
the
Riemann surface degenerates to a sphere above the phase transition point.

\section{Integrable structure of SUSY YM theories and the dispersion laws}

In this section we compare the data governing the vacuum structure of
SUSY YM theory and the one from the Peierls model.Low energy effective action
in YM theory is fixed by the Riemann surface and holomorphic differential
defined on it.Prepotential $\cal{F}$ can be derived from the relations
\be
\label{periods}
a_{D_{i}}=\frac{d\cal{F}}{da_{i}}\\
a_{i}= \oint_{A_{i}}\lambda \\
a_{D_{i}}=\oint_{B_{i}}\lambda   ;\lambda=dS=Edp
\ee
where $A_{i}$ and $B_{i}$ are the cycles on the curve $\Sigma$.Generically
the equation for the Toda curve of the length N reads \cite{toda}
\be
y+y^{-1}=P_{N}(E)
\ee
However it was proved \cite{dp} that only one gap vacuum configuration
is stable,therefore the generic polynomial denenerates.
In what follows we will treat both Riemann surface and dS in terms of the
quasiparticles.Namely $\Sigma$ coincides with the dispersion law
after the identification $y=exp(iNp(E))$ and
dS with the differential of the quasiparticle energy \footnote{note that
recently the dispersion laws of the quasiparticles were used for the
general classification of the low energy effective actions. \cite{Leut}}

Let us start with rather general arguments concerning the relation (15).
We adopt the following viewpoint;Toda chain reflecting Seiberg-Witten
solution is substituted by the system of fermions on the dynamical lattice
and the Lax operator for the Toda system is assumed to coincide with the
fermionic Hamiltonian.Therefore we would like to investigate the coupled
system of fermions on the lattice.The meaning of the variables $a_{i}$
as lattice action variables was established before \cite{toda}
and now we are going to
clarify the general meaning  of  (15) where according to
\cite{toda} $\cal{F}$ is nothing but the logariphm of the quasiclassical
("averaged") $\tau$ function.

If we compare (15) with the wellknown relation
\be
\Delta{\theta_{i}}=\frac{\partial<S(x_{i},I_{i})>}{\partial I_{i}}
\ee
where $ <S>$ is the averaged action and $\Delta{\theta_{i}}$ -so called
Hanney angle,
the immediate identification of the latter with
$a_{D_{i}}$ arises.Remind that the Hanney angle is the quasiclassical
analogue of the quantum Berry phase and  relation between them reads as
follows
\be
\Delta{\theta_{i}}=\frac{\partial\gamma}{\partial n_{i}} +O(h)
\ee
where $n_{i}$ appears in the Bohr-Sommerfeld quantization condition.It is
worth noting that usually the nontrivial Hanney angles comes from the
intersection of the isoenergetic surfaces where the auxiliary magnetic
charges are distributed and this is in the qualitative agreement with
the "magnetic" interpretation of $a_{D_{i}}$ in 4d case.Note that
nontriviality of the Hanney angle actually is dictated by the nontrivial
"bundle of the action-angle variables" over the parameter space.The latter
coincides now with the spectral curve.Note also that this remark is in
the perfect agreement with the definition of the Hanney angle as the
integral over the closed loop in the parameter space
\be
\Delta{\theta_{i}}=\int dB<pdq>
\ee
where B (coordinate on the spectral curve in  our case) is the
external parameter.One more argument supporting our point of view is
the convenient appearence of the Berry phase which is known to
have the meaning of the effective action arising after the averaging
over the fast degrees of freedom.Therefore the discussion  above suggests
that  low energy effective action describes the lattice degrees
of freedom after the averaging over the fermions indeed.

Let us now relate the parameters of the Peierls model with the ones
of 4d theory.At first it is useful to find the length of the lattice.For this
purpose remind that Toda chain can be derived from the Calogero-Moser system
via the dimensional transmutation proceedure.
This means that we add the one adjoint multiplet with the mass M to the
N=2 YM Lagrangian.Integrable system behind this theory is known to be the
Calogero-Moser one where M plays the role of the coupling constant g.
We can consider the following limiting proceedure \cite{cal}
\be
g=g_{0}e^{a}\\
g\to \infty ;g_{0} fixed
\ee
where Toda chain variables $\phi_{n}$
are defined in terms of the Calogero particles
$x_{n}$ as
\be
x_{n}=na +\phi_{n}
\ee

and a is the average distance between the cites.Therefore the
total length can be identified with the coupling constant
$\tau_{0}=\frac{4\pi i}{\alpha^{2}}+\frac{\theta}{2\pi}$
taken at the UV scale.To get the
interpretation of the another parameter of the Peierls model,namely the
density of the fermions let us compare the solution above (9) and the SW one.
Immediately one gets (we restrict ourselves by SU(2) case) for the fermionic
density $\rho=\frac{\mid \tau(u) \mid}{4}$ therefore the number of the
fermions in the continuum case  equals to the ratio of the renormalized and
UV coupling constants.For the discrete case the fermionic density can be read
off from (9).

\section{BPS states}

One of the most intriguing questions which has to be answered in this approach
is about the proper meaning of BPS states.We will show that it
can be interpreted as the string of quasiparticles described above on the
complexified Fermi surface.Let us remind that BPS states in N=2 YM theory
have masses (for SU(2) case)
\be
M_{n,m}=\mid na+ma_{D}\mid
\ee
so in the context of the Peierls model
this formulae asquires the meaning of the energy of the fermions
filling the allowed or forbidden zones.The "strings of the quasiparticles"
are wrapped around the
noncontractable cycles on the spectral surface and numbers n,m correspond
to the number of the strings.It is useful to compare our string picture of
BPS states with the one suggested in \cite{vafa} where these strings were
interpreted as the intersection of 2-branes which lie on 5-branes with
the Riemann surface $\Sigma$.It was shown that these strings are selfdual in
the sense of \cite{selfdual}.

Given a quasiparticle picture we can discuss the composite nature of the
selfdual strings.Indeed there are some conjectures about the composite
structure of string in the literature \cite{vafa,vafatop} and
the elementary object was assumed to be 0-brane living on p-brane.This
conjecture gets  some evidence in our approuch.Riemann surface
which is interpreted as a spectral curve for the integrable system
has another interpretation as a world-volume(or part of it) of D-branes.
Gauge $U(1)$ fields on the single D-brane with RR charge
\cite{polchinski} or SU(N) for N coinciding D-branes \cite{witbran} provide
the natural topological field theory \cite{vafatop} and therefore the
dynamical degrees of freedom in the Hitchin approach.Moreover the open
strings ending on the branes give rise to the Wilson line sources.With this
identification we can assume that our fermions are the 0-branes on the
spectral surface of the integrable system.

Let us emphasize that W-bosons and monopoles in SW solution also
have to be treated as the composite objects built from the
spectral fermions.

Now we present the argument that the system can undergo the temperature
phase transition which resembles the deconfainment phase transition in QCD.
The main point we would like to mention is that the mass gap analogous
to the $\Lambda_{QCD}$ disappeares above some critical temperature \cite{cp}.
In what
follows we will consider the simplest SU(2) case but all features are
captured already is this situation.

The question can be formulated as follows;at what  temperature
the equations determining the vacuum state  allow solution without
the forbidden zone?After the simple calculation one gets the equation
\be
\label{phtr}
8{\pi}T_{c}^{\frac{1}{2}}=\int\limits_{z_{1}}^\infty dz
\frac{thz}{z\sqrt{z-2z_{1}}} \\
2zT_{c}=E-\mu; 2z_{1}T_{c}=E_{1}-\mu
\ee
where $\mu$ is chemical potential,$E_{1}$ can be identified with the
vacuum expectation value of the scalar field in YM theory.If
$\mu -E_{1} \gg T_{c}$  the critical temperature appeares to be
\be
T_{c}\propto \mid E_{1}-\mu \mid exp(-4\pi \sqrt{\mid E_{1}-\mu\mid})
\ee
Thus we see that the theory manifests the phase transition behaviour and
the value of the critical temperature is proportional to
$\Lambda_{QCD}$ .

\section{Conclusions}

We have shown that one can formulate the quasiparticle picture behind
the integrable structure in N=2 YM theories.In such approach
Lax operator has the meaning of the Hamiltonian for the fermions
interacting with the lattice degrees of freedom.BPS states including
W-bosons and monopoles get
interpretation as the strings of the  quasiparticles.
Spectral surface of the integrable system plays the
role of the dispersion law for the quasiparticles. Moreover we provide
some evidence for the compositeness of the selfdual strings
which have been previously
identified with  BPS states in the membrane picture.
Let us emphasize that the Riemann surface defining
the ground state of the model appears in the dynamical way.

Let us make a few conjectures about the
possible nature of the fermions playing
the key role in  our consideration.To get some feeling let us
consider the effective mass of the quasiparticle.The velocity can be
found as $v=\frac{\partial E}{\partial p}$ and an easy calculation
shows that it vanishes at the branching points of the spectral curve,
therefore at these points quasiparticle asquires the infinite mass.The
appearence of the points where the quasiparticles have the infinite masses
indicates the possibility for the so called fermion condensate to
appear \cite{khodel} which is known in the solid state physics in
the strong coupling regime.If we conjecture that this phenomena and
the fermion condensate formation in the QFT have the related origin one can
look for the fermionic degrees of freedom condensing in  the SUSY vacuum.
This is known to happen with gluinos.One more suggestive picture
comes from the usual QCD where fermions develop the band energy structure
in the background of the instanton-antiinstanton ensemble which
results in the usual fermion condensate.Moreover one can speculate that
the wellknown potential in the Chern number coordinate becomes the dynamical
variable.To include into the game the matter degrees of  freedom one has
to allow for the fermions
the jumps between all lattice cites for the adjoint matter
\cite{cal} or to assign the spin degrees of  freedom to the lattice
for the fundamental matter \cite{chains}.
However we can,t present now more rigorous arguments and hope to
return to this point elsewhere.

\section{Acknowledgements}

I am indebted to I.Krichever,H.Leutwyler,A.Marshakov and N.Prokof,ev
for the useful  discussions.I thank H.Leutwyler for the hospitality
at the Institute for Theoretical Physics at Bern University where
the part of the work was done.

The work was supported by INTAS 1010--CT93--0023.


\bigskip

\begin{thebibliography}{12}

\bibitem{sw}
N.Seiberg and E.Witten, Nucl.Phys. {\bf B426} (1994) 19;
hepth/9407087\\
N.Seiberg and E.Witten, Nucl.Phys. {\bf B431} (1994) 484;
hepth/9408099

\bibitem{toda}
A.Gorsky, I.Krichever, A.Marshakov, A.Mironov and A.Morozov,
Phys.Lett. {\bf B355} (1995) 466; hepth/9505035\\
E.Martinec and N.Warner,Nucl.Phys.{\bf B459}(1996),97, hepth/9509161\\
T.Nakatsu and K.Takasaki,Mod.Phys.Lett.{\bf A11}(1996),157, hepth/9509162\\
T.Eguchi  and S.Yang, hep-th/9510183\\
E.Martinec and N.Warner, hepth/9511052\\
T.Nakatsu and K.Takasaki, hepth/9603069\\
I.Krichever,D.Phong,hepth/9604199 \\
C.Gomez,R.Hernandez,E.Lopez hepth/ 9604057

\bibitem{cal}
R.Donagi and E.Witten,Nucl.Phys.{\bf B460}(1996),299, hepth/9510101\\
E.Martinec,Phys.Lett.{\bf B367}(1996),91; hepth/9510204\\
A.Gorsky and A.Marshakov, hepth/9510224\\
H.Itoyama and A.Morozov, hepth/9511126,hepth/9512161


\bibitem{chains}
A.Gorsky, A.Marshakov, A.Mironov and A.Morozov,
hepth 9603140;hepth 9604078


\bibitem{mm} H.Itoyama and A.Morozov, hepth/9601168 \\
 A.Marshakov, hepth/9602005

\bibitem{cp} S.Brazovskii,JETP {\bf 78}(1980),678 \\
 E.Belokolos,Teor.i Mat.Phys. {\bf45}(1980),268\\
S.Brazovskii,I.Dzyaloshinskii,N.Kirova,JETP {\bf 81}(1981),2279

\bibitem{dp}
S.Brazovskii,I.Dzyaloshinskii,I.Krichever,JETP {\bf 83}(1982),389\\
I.Dzyaloshinskii,I.Krichever,JETP,{\bf 83}(1982),1576\\
I.Krichever,Func.Anal.Appl.,{\bf 20}(1986),42\\
I.Dzyaloshinskii,I.Krichever,Ya.Khronek,JETP,{\bf 94}(1988),344


\bibitem{vafa} A.Klemm,W.Lerche,P.Mayr,C.Vafa,N.Warner,hepth/ 9604034\\
M.Douglas,M.Li hepth/ 9604041

\bibitem{polchinski}
J.Polchinski,hepth /9507017

\bibitem{selfdual}
E.Witten,hepth/ 9507121\\
A.Strominger,hepth/ 9512059  \\
O.Hanor,A.Hanany,hepth/ 9602120

\bibitem{witbran}
E.Witten,hepth 9511030


\bibitem{vafatop}
C.Vafa,hepth/9511088\\
M.Bershadsky,V.Sadov,C.Vafa, hepth/9511222

\bibitem{Leut} H.Leutwyler,Phys.Rev.{\bf 49}(1994),3033

\bibitem{khodel}
V.Khodel,V.Shaginyan,JETP.Lett.{\bf 51}(1990),488


\end{thebibliography}
\end{document}